\documentclass{article}

\usepackage[tbtags]{amsmath}
\usepackage{amsfonts,amssymb,amsthm}
\usepackage{mathrsfs}

\theoremstyle{plain}
\newtheorem{THEOREM}{Theorem}

\newtheorem{COR}{Corollary}

\theoremstyle{definition}
\newtheorem{REMARK}{Remark}
\newtheorem{demo}{Proof}

\def\qed{\quad\text{$\square$}}

\begin{document}

\title{On the Dispersion Law of the Form
$\varepsilon(p)=\hbar^2p^2/2m+\widetilde V(p)-\widetilde V(0)$
for Elementary Excitations of a Nonideal Fermi Gas
in the Pair Interaction Approximation
$(i\leftrightarrow j)$, $V(|x_i-x_j|)$}

\author{V.~P.~Maslov\thanks{Moscow State University,
v.p.maslov@mail.ru}}

\date{}

\maketitle

\markboth{Maslov}{Dispersion Law for Elementary Excitations of a Nonideal Fermi Gas}

\begin{abstract}
 We give a derivation of the dispersion law
$\varepsilon(p)=\hbar^2p^2/2m
+\widetilde V(p)-\widetilde V(0)$,
where~$\widetilde V(p)$
is the Fourier transform of the pair
interaction potential~$V(r)$.
(The interaction between particles~$x_i$
and~$x_j$
is~$V(|x_i-x_j|)$.)

\end{abstract}

\begin{flushright}
{\vskip -5mm
\small\it Dedicated to the memory of my dear student, physicist
\\
 Vladimir Vladimirovich Belov}
\end{flushright}

\section{INTRODUCTION}
\label{ss1}

 In the famous paper~\cite{1},
Bogolyubov obtained the spectrum (energy levels)
of a nonideal Bose--Einstein gas.

 The formula for the spectrum reads
\begin{equation}
\label{eq1}
\lambda_{p}=-\hbar^2vp+\varepsilon(p),\qquad
\varepsilon(p)=\sqrt{\biggl(\frac{\hbar^2p^2}{2m}+\widetilde{V}(p)\biggr)^2
-\widetilde{V}^2(p)}\,.
\end{equation}
where~$v$ is the fluid (gas) velocity
and~$\widetilde V(p)$ is the Fourier transform
of the pair interaction potential~$V(|x_i-x_j|)$
for particles~$i$
and~$j$.

 For small~$p$,
the term~$\varepsilon(p)$
behaves as
$|p|\sqrt{\widetilde V(0)}$,
which corresponds to sound (phonons),
provided that~$\widetilde V(0)>0$.
 The last condition can be rewritten as
$\int_0^\infty V(r)\,dr$
and means that the repulsion for small~$r$
exceeds the attraction occurring in helium
for large~$r$.
 The Fourier transform~$\widetilde V(p)$
tends to zero
as~$p\to\infty$,
and the leading term is of the order of~$p^2/2$.
 The point~$p_0$ of minimum
of the radicand, where the~$p$-derivative
vanishes, is called the roton part of the
spectrum~$\varepsilon(p)$.

 In the same journal issue
where Bogolyubov's paper was published,
Landau gave a general argument
based on experiments and showing that the
curve~$\varepsilon(p)$ should have exactly this
general form, i.e., be linear near~$p=0$,
have one point of maximum and then a point of minimum,
and then tend to infinity.

 This curve is called the Landau curve
(the~1962 Nobel prize).

 The superfluidity of helium-4, which is a Bose gas,
was discovered by Kapitsa in~1930 (the~1978
Nobel prize).

 However, while the superfluid state
in the Bose case is caused by short-range repulsion,
it is known that the superfluid state has to be caused
by long-range attraction, which enables
the formation of Cooper pairs (the~1972
Nobel prize).

 The superfluidity of a Fermi fluid (helium-3)
was discovered experimentally in the early 1970s
by the American scientists
David Lee, Douglas Osheroff, and Robert Richardson
(the~1996 Nobel prize).

 Landau wrote a year after publishing his result:
``Recently, Bogolyubov has succeeded in determining
the energy spectrum of a Bose--Einstein gas with weak
interaction between particles in the general case by
an ingenious application
of second quantization''~\cite[p.~43]{2}.

 Specifically, Bogolyubov used
Dirac's idea: since the number of particles
is large, one can deem the commutator of
the creation and annihilation operators in the Bose
statistics to be small.
 The so-called Bogolyubov asymptotic $(u,v)$-transform
leads to the same result.

In the Fermi statistics, one considers the anticommutator of
the creation--annihilation operators, so that the above
argument fails to provide passage to~``$c$-numbers.''
 This passage is only possible for pairs of fermions.

 The theory of ultrasecond  quantization developed by
the present author permits one to obtain an even simpler
formula, based on long-range attraction,
for a Fermi gas.
 Namely,
\begin{equation}
\label{eq2}
\lambda_{p}=-\hbar^2pv+\biggl|
\frac{\hbar^2p^2}{2m}+\widetilde{V}(p)
-\widetilde{V}(0)\biggr|.
\end{equation}
This formula is the subject
of the present paper.

 The Landau curve also arises
in this case, and the main role is played
by the long-range attraction provided by the
interaction potential.
 To show this, we first study the behavior
of~$\widetilde V(p)$.

\begin{REMARK}
The second-quantized Schr\"odinger equation
\textit{identically}
coincides with the~$N$-particle Schr\"odinger equation
up to a unitary transformation.
 We mean that the second-quantized equation
is a convenient representation
of the~$N$-dimensional system of Schr\"odinger equations.
 This representation (like the $p$-representation,
the interaction representation, etc.)
is ``identical'' up to a unitary transformation.
 It contains no approximations and is, in this sense,
an identity transformation.
 In just the same way, ultrasecond quantization,
introduced by the author, coincides up to a unitary
transformation with the~$N$-particle
Schr\"odinger equation for symmetric and antisymmetric
solutions.
 For some reason, this new identity scares
off theoretical physicists so much that they begin
to cross themselves, saying ``keep away from me,''
and publicly burn the author's books as well as his likeness.
\end{REMARK}

 However, Bogolyubov's asymptotic
$(u,v)$-transform, which is valid for bosons,
coincides in the fermionic case with the ultrasecond
quantization method only for Bardeen's
interaction potential.
 This transformation is approximate rather than exact
and gives a wrong pair interaction asymptotics.
 Apparently, it is only after
a trap experiment
with a Fermi gas, similar to the one
carried out in the remarkable paper~\cite{3},
is conducted for the ultrasecond
quantization method
that theoretical physicists will believe in the method.

\section{``GENERAL POSITION'' THEOREM}
\label{ss2}
 In mathematics, the notion of ``general position''
for points of a surface, say, with respect to
projection onto a coordinate plane, is well known and
was in particular used in an essential way by Arnold, Varchenko,
and Gusein-Zade.
 In physics, this special case of general position means
in particular that the optical focus of a lens refracting
light rays so that they all converge at a same point is not
a generic situation---once the lens is moved slightly,
the focus spreads into a small light spot formed by a network
of intersecting caustics.

 On the plane, a perpendicular drawn to
the abscissa axis is not in general
position---after a small rotation of the coordinate
axes, this line is no longer projected into a
single point on the abscissa axis.
 If the first two derivatives are zero
at a point of a smooth curve, then the curve
is not in general position with respect to a
small rotation of the axes.
 One can rotate the axes in such a way that, in a
majority of cases, the smooth curve will not have
such points.
 Then one speaks of
a curve in ``general position.''

 As to a point where the derivative is zero,
a small rotation destroys this property.
 Then one speaks of a ``generic'' point of
a smooth curve.
 The derivative at a generic point of a smooth curve
is nonzero.
 In particular, the derivative
of a generic smooth curve
on the half-line is nonzero at the origin.
 If this is not the case, then a small rotation
of the coordinate axes destroys this property.
 Thus, a small rotation does not affect the property
of the derivative to be nonzero at the origin.
 The following elementary theorem holds.

\begin{THEOREM}
Let the original pair interaction potential~$U(x-y)$,
$x,y\in\nobreak\mathbb R^3$,
be radially symmetric and equal to~$V(r)$,
where
$r=\sqrt{\sum_{i=1}^3(x_i-y_i)^2}$
and~$\lim_{r\to0}V(r)r\ne\nobreak0$.
 Suppose that the three-dimensional Fourier transform
$\widetilde V(|p|)$,
$|p|\ge0$,
is in general position at the origin
$|p|=0$.
 Then
\begin{equation}
\label{eq3}
\lim_{p\to\infty}p^2\widetilde{V}(p)\ne0,\qquad
\lim_{r\to\infty}r^4V(r)\ne0.
\end{equation}
\end{THEOREM}

\begin{demo}
 Since~$V(\xi)$
is spherically symmetric, it follows that
so is the three-dimensional
Fourier transform~$\widetilde V(\xi)$.

 By passing to the spherical coordinates,
we obtain the relation
\begin{equation}
\label{eq4}
 V(r)=\frac1r\int^\infty_{-\infty}
\widetilde{V}(\xi)\sin(\xi r)\xi\,d\xi.
\end{equation}

 Since~$\widetilde V(|\xi|)$
tends to infinity at the rate of~$\mathrm{const}/|\xi|^2$,
it follows that
$$
V(r)=\frac{\mathrm{const}}r+O(1)
$$
as $r\to0$. The converse is also true.

 The condition of general position
implies that
$$
\widetilde V'(0)\ne0.
$$
 One can readily see that
\begin{equation}
\label{eq5}
\widetilde{V}'(0)=-\frac{1}{4\pi}
\lim_{r\to\infty}r^4 V(r).
\end{equation}

 Indeed, by replacing the integral~\eqref{eq4}
by the integral over the finite interval~$[-A,A]$,
where~$A$
is sufficiently large, by making the change of variables
$\xi r=\eta$,
by integrating by parts, and by passing to the limit
as~$r\to\infty$ and~$A\to\infty$,
we arrive at the assertion of the theorem.
\qed \end{demo}

 In fact, the interaction potential is not
exactly spherically symmetric in view of the
complicated structure of the molecule.
 In principle, one should take into account
the influence of all particles forming the
molecule.
 If, nevertheless, we treat the molecule
as a whole and approximate the interaction
potential by a spherically symmetric function,
we should at least assume that the potential
is ``generic.''

\begin{COR}
 One has
\begin{equation}
\label{eq6}
\widetilde{V}(p)-\widetilde{V}(0)
\sim-4\pi|p|\lim_{r\to\infty}r^4V(r) .
\end{equation}
as~$|p|\to0$.
Since the potential~$V(r)$
is attractive as~$r\to\infty$,
it follows that
\begin{equation}
\label{eq7}
-4\pi\lim_{r\to\infty}r^4 V(r)=c>0.
\end{equation}
\end{COR}

 As~$p\to\infty$,
the expression under the modulus sign in~\eqref{eq2}
tends to infinity at the rate of~$\hbar^2p^2/2m$,
since
$\widetilde V(p)\to\infty$
at the rate of~$1/|p|^2$.

 Thus, we see that formula~\eqref{eq2} gives the Landau
curve for a Fermi fluid.

\begin{REMARK}
 Consider the Lennard--Jones potential and add
a small smooth radially symmetric potential to it.
 The probability that the new potential still
decays at the rate of~$1/r^6$
is zero with respect to the entire set of potentials.
 With probability~$1$, the decay will be
at the rate of~$1/r^4$.
 There is no known physical law that
would prohibit this, and the existence
of such a law is unlikely.
\end{REMARK}

\section{ULTRASECOND QUANTIZATION}
\label{ss3}
 Ultrasecond quantization for problems
in quantum mechanics and statistical physics
was introduced in~\cite{4}--\cite{7}.
 Recall the notation and main facts for the case
of quantization with respect to ``particle--number'' pairs
and with respect to pairs of particles.
 Quantization with regard to pairs will permit us
to take into account pair correlations of particles
when constructing the asymptotics.
 The ultrasecond
quantization space is the bosonic
Fock space~$\mathscr F$~\cite{8}.
 Let~$\widehat b^+(x,s)$ be the creation operator
for particles with number~$s$, and
let~$\widehat b^-(x,s)$ be the annihilation operator
for particles with number~$s$
in~$\mathscr F$~\cite{8}.
  Next, let~$\widehat B^+(x,x')$ be the creation
operator for a pair of particles, and
let~$\widehat B^-(x,x')$ be the annihilation
operator for a pair of particles in this space.
 These operator satisfy the fermionic
anticommutation relations
\begin{equation}
\begin{gathered}
\label{eq8}
\{\widehat{b}^-(x,s),\widehat{b}^+(x',s')\}
=\delta_{ss'}\delta(x-x'),\qquad
\{\widehat{b}^{\pm}(x,s),\widehat{b}^{\pm}(x',s')\}=0,
\\
[\widehat{B}^{-}(x_1,x_2)\widehat{B}^{+}(x'_1,x'_2)]
=\delta(x_1-x'_1)\delta(x_2-x'_2),\qquad
[\widehat{B}^\pm(x_1,x_2),\widehat{B}^\pm(x'_1,x'_2)]=0,
\\
[\mspace{2mu}\widehat{b}^\pm(x,s),\widehat{B}^\pm(x'_1,x'_2)]
=[\mspace{2mu}\widehat{b}^\pm(x,s),\widehat{B}^\mp(x'_1,x'_2)]=0.
\end{gathered}
\end{equation}
 For bosons,
$b^-$
and~$b^+$
satisfy similar commutation relations.

 Next,
$\Phi_0$ is the vacuum vector
in~$\mathscr F$ with the following properties:
\begin{equation}
\label{eq9}
\widehat{b}^-(x,s)\Phi_0=0,\qquad
\widehat{B}^-(x_1,x_2)\Phi_0=0.
\end{equation}
 The variable~$x$
ranges over the three-dimensional
torus~$L_1\times L_2\times L_3$,
which will be denoted by~$\mathbf T$.
 The variable
$s=0,1,\dots$ is discrete;
it is called the \textit{number}, or the
\textit{statistical spin}.
 Each vector~$\Phi\in \mathscr F$
can be uniquely represented in the form
\begin{equation}
\begin{split}
\label{eq10}
\Phi&=\sum^{\infty}_{k=0}\sum^{\infty}_{M=0}
\frac1{k!\mspace{1mu}M!}\sum^{\infty}_{s_1=0}\dots
\sum^{\infty}_{s_k=0}\int\dots\int
dx_1\dots dx_k\,dy_1\dots dy_{2M}
\\
&\qquad\qquad
\times\Phi_{k,M}(x_1,s_1;\dots;x_k,s_k;
y_1,y_2;\dots;y_{2M-1},y_{2M})
\\
&\qquad\qquad
\times\widehat{b}^+(x_1,s_1)\dotsb
\widehat{b}^+(x_k,s_k)\widehat{B}^+(y_1,y_2)\dotsb
\widehat{B}^+(y_{2M-1},y_{2M})\Phi_0,
\end{split}
\end{equation}
where the function
$\Phi_{k,M}(x_1,s_1;\dots;x_k,s_k;
y_1,y_2;\dots;y_{2M-1},y_{2M})$
is symmetric with respect to transpositions
of the pairs~$(x_j,s_j)$
and~$(x_i,s_i)$
as well as of the pairs
$(y_{2j-1},y_{2j})$
and~$(y_{2i-1},y_{2i})$.
 In the bosonic case, one introduces
the subspace~$\mathscr F^{\mathrm{Symm}}_{k,M}$
of vectors~$\Phi$
such that~$\Phi_{k',M'}=0$
for
$(k',M')\ne(k,M)$ and~$\Phi_{k,M}$
is a symmetric function of the variables
$x_1,x_2,\dots,x_k,y_1,y_2,\dots,\allowbreak y_{2M}$.
 In the fermionic case, one in a similar way
introduces the subspace~$\mathscr F^{\mathrm{Asymm}}_{k,M}$
of vectors~$\Phi$
such that~$\Phi_{k',M'}=0$
for~$(k',M')\ne(k,M)$
and~$\Phi_{k,M}$ is an antisymmetric function
of the variables
$x_1,x_2,\dots,x_k,y_1,y_2,\dots,y_{2M}$.
 The orthogonal projection
onto~$\mathscr F^{\mathrm{Symm}}_{k,M}$
in~$\mathscr F$
has the form~\cite{4}--\cite{7}
\begin{equation}
\begin{split}
\label{eq11}
\widehat{\Pi}^{\mathrm{Symm}}_{k,M}&=\frac1{k!\mspace{1mu}M!}
\sum^{\infty}_{s_1=0}\dots\sum^{\infty}_{s_k=0}
\int\dots\int dx_1\dots dx_k\,dy_1\dots dy_{2M}
\\
&\qquad\qquad
\times\widehat{b}^+(x_1,s_1)\dotsb
\widehat{b}^+(x_k,s_k)\widehat{B}^+(y_1,y_2)\dotsb
\widehat{B}^+(y_{2M-1},y_{2M})
\\
&\qquad\qquad
\times\operatorname{Symm}_{x_1\dots x_ky_1\dots y_{2M}}
\bigl(\mspace{2mu}\widehat{b}^-(x_1,s_1)\dotsb\widehat{b}^-(x_k,s_k)
\\
&\qquad\qquad\qquad\qquad
\times\widehat{B}^-(y_1,y_2)\dotsb\widehat{B}^-(y_{2M-1},y_{2M})
\bigr)
\\
&\qquad\qquad
\times\exp\biggl(-\sum^{\infty}_{s=0}\int dx\,\,
\widehat{b}^+(x,s)\widehat{b}^-(x,s)
-\iint dy\,dy'\,\,\widehat{B}^+(y,y')
\widehat{B}^-(y,y')\biggr),
\end{split}
\end{equation}
where
$\operatorname{Symm}_{x_1\dots x_ky_1\dots y_{2M}}$
is the operator of symmetrization
with respect to the variables
$x_1,\dots,x_k,y_1,\dots,\allowbreak y_{2M}$
and the operators~$\widehat b^+(x,s)$,
$\widehat b^-(x,s)$,
$\widehat B^+(y,y')$,
and~$\widehat B^-(y,y')$
are Wick order~\cite{8}.
 The orthogonal projection
onto~$\mathscr F^{\mathrm{Asymm}}_{k,M}$
in~$\mathscr F$
has the form~\cite{4}
{\allowdisplaybreaks
\begin{align*}
\widehat{\Pi}^{\mathrm{Asymm}}_{k,M}
&=\frac1{k!\mspace{1mu}M!}
\sum^{\infty}_{s_1=0}\dots\sum^{\infty}_{s_k=0}
\int\dots\int dx_1\dots dx_k\,dy_1\dots dy_{2M}
\\
&\qquad\qquad
\times\widehat{b}^+(x_1,s_1)\dotsb\widehat{b}^+(x_k,s_k)
\widehat{B}^+(y_1,y_2)\dotsb\widehat{B}^+(y_{2M-1},y_{2M})
\\
&\qquad\qquad
\times\operatorname{Asymm}_{x_1\dots x_ky_1\dots y_{2M}}
\bigl(\mspace{2mu}\widehat{b}^-(x_1,s_1)\dotsb\widehat{b}^-(x_k,s_k)
\\
&\qquad\qquad\qquad\qquad
\times\widehat{B}^-(y_1,y_2)\dotsb\widehat{B}^-(y_{2M-1},y_{2M})
\bigr)
\\
&\qquad\qquad
\times\exp\biggl(-\sum^{\infty}_{s=0}\int dx\,\,
\widehat{b}^+(x,s)\widehat{b}^-(x,s)-\iint dy\,dy'\,\,
\widehat{B}^+(y,y')\widehat{B}^-(y,y')\biggr),
\end{align*}}
where
$\operatorname{Asymm}_{x_1\dots x_ky_1\dots y_{2M}}$
is the operator of antisymmetrization
with respect to the variables
$x_1,\dots,x_k,\allowbreak y_1,\dots,y_{2M}$.
 From now on, the operators~$\widehat b^+(x,s)$,
$\widehat b^-(x,s)$,
$\widehat B^+(y,y')$,
and~$\widehat B^-(y,y')$
are Wick ordered unless specified otherwise.

 In what follows, we consider a system
of~$N$ identical particles on the torus~$\mathbf T$.
 We assume that the Hamiltonian
of~$N$
bosons or fermions has the form
\begin{equation}
\label{eq12}
\widehat{H}_N=-\frac{\hbar^2}{2m}\sum^{N}_{j=1}\Delta_j
+\sum^{N}_{j=1}\sum^{N}_{l=j+1}V(x_j-x_l).
\end{equation}
 According to~\cite{4}, the ultrasecond-quantized
Hamiltonian corresponding to this operator
in the bosonic case has the form
\begin{equation}
\begin{split}
\label{eq13}
\overline{\widehat{H}}_B
&=\sum^{\infty}_{k=0}\sum^{\infty}_{M=0}\frac1{k!\mspace{1mu}M!}
\sum^{\infty}_{s_1=0}\dots\sum^{\infty}_{s_k=0}
\int\dots\int dx_1\dots dx_k\,dy_1\dots dy_{2M}
\\
&\qquad\qquad
\times\widehat{b}^+(x_1,s_1)\dotsb\widehat{b}^+(x_k,s_k)
\widehat{B}^+(y_1,y_2)\dotsb\widehat{B}^+(y_{2M-1},y_{2M})
\widehat{H}_{k+2M}
\\
&\qquad\qquad
\times\operatorname{Symm}_{x_1\dots x_ky_1\dots y_{2M}}
\bigl(\mspace{2mu}\widehat{b}^-(x_1,s_1)\dotsb\widehat{b}^-(x_k,s_k)
\\
&\qquad\qquad\qquad\qquad
\times\widehat{B}^-(y_1,y_2)\dotsb\widehat{B}^-(y_{2M-1},y_{2M})
\bigr)
\\
&\qquad\qquad
\times\exp\biggl(-\sum^{\infty}_{s=0}\int dx\,\,
\widehat{b}^+(x,s)\widehat{b}^-(x,s)-\iint dy\,dy'\,\,
\widehat{B}^+(y,y')\widehat{B}^-(y,y')\biggr).
\end{split}
\end{equation}
In the fermionic case, the corresponding
operator~$\overline{\widehat H}_F$
is expressed by a similar formula
with~$\operatorname{Symm}$
replaced by~$\operatorname{Asymm}$.
 By analogy with~\eqref{eq12} and~\eqref{eq13},
we assign an ultrasecond-quantized
operator~$\overline{\widehat A}$~\cite{4}
to every~$N$-particle operator
$$
\widehat A_N\biggl(\stackrel2{x_1},\dots,\stackrel2{x_N};
-i\stackrel1{\frac\partial{\partial x_1}}\mspace{1mu},\dots,
-i\stackrel1{\frac\partial{\partial x_N}}\biggr).
$$
 For example, the ultrasecond-quantized
identity operator corresponding to the identity operator
in the bosonic case has the form
{\allowdisplaybreaks
\begin{align}
\overline{\widehat{E}}_B
&=\sum^{\infty}_{k=0}\sum^{\infty}_{M=0}\frac1{k!\mspace{1mu}M!}
\sum^{\infty}_{s_1=0}\dots\sum^{\infty}_{s_k=0}
\int\dots\int dx_1\dots dx_k\,dy_1\dots dy_{2M}
\nonumber
\\
&\qquad\qquad
\times\widehat{b}^+(x_1,s_1)\dotsb\widehat{b}^+(x_k,s_k)
\widehat{B}^+(y_1,y_2)\dotsb\widehat{B}^+(y_{2M-1},y_{2M})
\nonumber
\\
&\qquad\qquad
\times\operatorname{Symm}_{x_1\dots x_ky_1\dots y_{2M}}
\bigl(\mspace{2mu}\widehat{b}^-(x_1,s_1)\dotsb\widehat{b}^-(x_k,s_k)
\nonumber
\\
&\qquad\qquad\qquad\qquad
\times\widehat{B}^-(y_1,y_2)\dotsb\widehat{B}^-(y_{2M-1},y_{2M})
\bigr)
\nonumber
\\
\label{eq14}
&\qquad\qquad
\times\exp\biggl(-\sum^{\infty}_{s=0}\int dx\,\,
\widehat{b}^+(x,s)\widehat{b}^-(x,s)-\iint dy\,dy'\,\,
\widehat{B}^+(y,y')\widehat{B}^-(y,y')\biggr),
\end{align}}
which is the sum of the projections~\eqref{eq11}.
 Likewise, the ultrasecond-quantized
identity operator in the fermionic case
is given by
$$
\overline{\widehat E}_F
=\sum^\infty_{k=0}\sum^\infty_{M=0}
\widehat\Pi^{\mathrm{Asymm}}_{k,M},
$$
and~$\operatorname{Symm}$ in~\eqref{eq14}
is replaced by~$\operatorname{Asymm}$.

 Consider the eigenvalue problem
\begin{equation}
\label{eq15}
\overline{\widehat{H}}_{B,F}\Phi
=\lambda\overline{\widehat{E}}_B\Phi,\qquad
\overline{\widehat{E}}\Phi\ne0,
\end{equation}
in the bosonic and fermionic cases.
 The following assertion was proved in~\cite{4}.
\textit{On the subspaces~$\mathscr F^{\mathrm{Symm}}_{k,M}$
and~$\mathscr F^{\mathrm{Asymm}}_{k,M}$
of the space~$\mathscr F$,
the operators~$\overline{\widehat H}_B$
and~$\overline{\widehat H}_F$,
respectively, acted upon by the projection
onto the~$N$-particle space
(of the operator~\eqref{eq21} of the number
of particles) coincide with the
operator~$\widehat H_N$}.
 Hence the eigenvalues~$\lambda$
of problem~\eqref{eq15} in the bosonic and fermionic
cases coincide with the corresponding
eigenvalues of the operators
$\widehat H_N$~\eqref{eq12}.
 If the commutators of the
operators~$\widehat b^-(x,s)$
and~$\widehat b^+(x,s)$,
as well as of~$\widehat B^-(x,y)$
and~$\widehat B^+(x,y)$,
are small of the order
of~$1/N$, then, according to~\cite{4},
the asymptotics of solutions of problem~\eqref{eq15}
is determined by the symbol corresponding to problem~\eqref{eq15}.
 In the bosonic case,
the pseudosymbol is defined in the form
{\allowdisplaybreaks
\begin{align}
&\mathscr H_B[b^*(\,\cdot\,),b(\,\cdot\,),
 B^*(\,\cdot\,),B(\,\cdot\,)]
\nonumber
\\
&\qquad
=\biggl\{\sum^{\infty}_{k,M=0}\frac1{k!\mspace{1mu}M!}
\sum^{\infty}_{s_1=0}\dots\sum^{\infty}_{s_k=0}
\int\dots\int dx_1\dots dx_k\,dy_1\dots dy_{2M}
\nonumber
\\
&\qquad\qquad\qquad\qquad\qquad
\times b^*(x_1,s_1)\dotsb b^*(x_k,s_k)
 B^*(y_1,y_2)\dotsb B^*(y_{2M-1},y_{2M})H_{k+2M}
\nonumber
\\
&\qquad\qquad\qquad\qquad\qquad
\times\operatorname{Symm}_{x_1\dots x_ky_1\dots y_{2M}}
\bigl(b(x_1,s_1)\dotsb b(x_k,s_k)
\nonumber
\\
&\qquad\qquad\qquad\qquad\qquad\qquad\qquad
\times B(y_1,y_2)\dotsb B(y_{2M-1},y_{2M})\bigr)\biggr\}
\nonumber
\\
&\qquad\qquad\qquad
\times\biggl\{\sum^{\infty}_{k',M'=0}\frac1{k'!\mspace{1mu}M'!}
\sum^{\infty}_{s'_1=0}\dots\sum^{\infty}_{s'_{k'}=0}
\int\dots\int dx'_1\dots dx'_{k'}\,dy'_1\dots dy'_{2M'}
\nonumber
\\
&\qquad\qquad\qquad\qquad\qquad
\times b^*(x'_1,s'_1)\dotsb b^*(x'_{k'},s'_{k'})
 B^*(y'_1,y'_2)\dotsb B^*(y'_{2M'-1},y'_{2M'})
\nonumber
\\
&\qquad\qquad\qquad\qquad\qquad
\times\operatorname{Symm}_{x'_1\dots x'_{k'}
y'_1\dots y'_{2M'}}
\bigl(b(x'_1,s'_1)\dotsb b(x'_{k'},s'_{k'})
\nonumber
\\
\label{eq16}
&\qquad\qquad\qquad\qquad\qquad\qquad\qquad
\times B(y'_1,y'_2)\dotsb B(y'_{2M'-1},y'_{2M'})\bigr)\biggr\}.
\end{align}}
 The expression for the symbol in the fermionic case
is similar except that~$\operatorname{Symm}$
in~\eqref{eq16}
is replaced by~$\operatorname{Asymm}$.
 The following identity holds for the
pseudosymbol~\eqref{eq16} in the bosonic case:
\begin{equation}
\label{eq17}
\mathscr H_B[b^*(\,\cdot\,),b(\,\cdot\,),
 B^*(\,\cdot\,),B(\,\cdot\,)]
=\frac{\operatorname{Sp}(\widehat\rho_B\widehat{H})}
{\operatorname{Sp}(\widehat\rho_B)}\mspace{1mu},
\end{equation}
where~$\widehat H$ and~$\widehat\rho_B$
are the second quantized operators,
\begin{equation}
\label{eq18}
\widehat{H}=\int dx\,\,\widehat{\psi}^+(x)
\biggl(-\frac{\hbar^2}{2m}\Delta\biggr)
\widehat{\psi}^-(x)+\frac12\iint dx\,dy\,\,V(x,y)
\widehat{\psi}^+(y)\widehat{\psi}^+(x)
\widehat{\psi}^-(y)\widehat{\psi}^-(x).
\end{equation}
 Here~$\widehat\rho_B$
depends on the functions~$b(x,s)$ and~$B(y,y')$:
\begin{equation}
\begin{split}
\label{eq19}
\widehat{\rho}_B&=\sum^{\infty}_{k=0}\sum^{\infty}_{M=0}
\frac1{k!\mspace{1mu}M!\mspace{1mu}(k+2M)!}\biggl(\sum^{\infty}_{s=0}\iint dx\,dx'\,\,b(x,s)b^*(x',s)
\widehat{\psi}^+(x)\widehat{\psi}^-(x')\biggr)^k
\\
&\qquad\qquad
\times\biggl(\iint dy_1\,dy_2\,\,B(y_1,y_2)
\widehat{\psi}^+(y_1)\widehat{\psi}^+(y_2)\biggr)^M
\\
&\qquad\qquad
\times\biggl(\iint dy'_1\,dy'_2\,\,B(y'_1,y'_2)
\widehat{\psi}^-(y'_1)\widehat{\psi}^-(y'_2)\biggr)^M
\exp\biggl(-\int dz\,\,
\widehat{\psi}^+(z)\widehat{\psi}^-(z)\biggr),
\end{split}
\end{equation}
where~$\widehat\psi^+(x)$ and~$\widehat\psi^-(x)$
are Wick ordered bosonic creation--annihilation
operators~\cite{7}.
 In the fermionic case,
one has the similar identity
$$
\mathscr H_F[b^*(\,\cdot\,),b(\,\cdot\,),
B^*(\,\cdot\,),B(\,\cdot\,)]
=\frac{\operatorname{Sp}(\widehat\rho_F\widehat H)}
{\operatorname{Sp}(\widehat\rho_F)}\mspace{1mu},
$$
where~$\widehat H$ and~$\widehat\rho_F$
are the following second-quantized operators:
\begin{align}
\widehat H&=\int dx\,\,\widehat\psi^+(x)
\biggl(-\frac{\hbar^2}{2m}\mspace{1mu}\Delta\biggr)
\widehat\psi^-(x)+\frac12\iint dx\,dy\,\,V(x,y)
\widehat\psi^+(x)\widehat\psi^+(y)
\widehat\psi^-(y)\widehat\psi^-(x)
\nonumber
\\
\widehat{\rho}_F&=\sum^{\infty}_{k=0}\sum^{\infty}_{M=0}
\frac1{k!\mspace{1mu}M!\mspace{1mu}(k+2M)!}\biggl(\iint dy_1\,dy_2\,\,B(y_1,y_2)\widehat{\psi}^+(y_1)
\widehat{\psi}^+(y_2)\bigg)^M
\nonumber
\\
&\qquad\qquad
\times\sum^{\infty}_{s_1=0}\dots\sum^{\infty}_{s_k=0}
\int\dots\int dx_1\,dx'_1\dots dx_k\,dx'_k
\nonumber
\\
&\qquad\qquad
\times b(x_1,s_1)b^*(x'_1,s_1)\dotsb
b(x_k,s_k)b^*(x'_k,s_k)
\nonumber
\\
&\qquad\qquad
\times\widehat{\psi}^+(x_1)\dotsb\widehat{\psi}^+(x_k)
\widehat{P}_0\widehat{\psi}^-(x'_k)\dotsb
\widehat{\psi}^-(x'_1)
\nonumber
\\
\label{eq20}
&\qquad\qquad
\times\biggl(\iint dy'_1\,dy'_2\,\,B(y'_1,y'_2)
\widehat{\psi}^-(y'_2)\widehat{\psi}^-(y'_1)\biggr)^M.
\end{align}
 Here~$\widehat\psi^+(x)$ and~$\widehat\psi^-(x)$
are the fermionic creation--annihilation
operators, and~$\widehat P_0$
is the projection
onto the vacuum vector of the fermionic Fock space.
 In general, the pseudosymbol of the ultrasecond-quantized
operator $\overline{\widehat L}$ corresponding to an arbitrary
second-quantized operator~$\widetilde L$
is expressed by the formula~\cite{4}
$$
L_{B,F}[b^*(\,\cdot\,),b(\,\cdot\,),
B^*(\,\cdot\,),B(\,\cdot\,)]
=\frac{\operatorname{Sp}(\widehat\rho_{B,F}\widehat A)}
{\operatorname{Sp}(\widehat\rho_{B,F})}\mspace{1mu}.
$$
 In the space~$\mathscr F$,
one introduces the ultrasecond-quantized
particle number operators~\cite{4}
\begin{equation}
\label{eq21}
\overline{\widehat{N}}_B
=\sum^{\infty}_{k=0}\sum^{\infty}_{M=0}(k+2M)
\widehat{\Pi}^{\mathrm{Symm}}_{k,M},\qquad
\overline{\widehat{N}}_F=\sum^{\infty}_{k=0}
\sum^{\infty}_{M=0}(k+2M)
\widehat{\Pi}^{\mathrm{Asymm}}_{k,M}.
\end{equation}
 Accordingly, the pseudosymbol
of the operator~$\overline{\widehat N}_B$
in the bosonic case has the form
\begin{equation}
\begin{split}
\label{eq22}
 N_B&=\biggl\{\sum^{\infty}_{k=0}\sum^{\infty}_{M=0}
\frac{k+2M}{k!\mspace{1mu}M!}\sum^{\infty}_{s_1=0}\dots
\sum^{\infty}_{s_k=0}\int\dots\int dx_1\,\dots dx_{k+2M}
\\
&\qquad\qquad
\times b^*(x_1,s_1)\dotsb b^*(x_k,s_k)
 B^*(x_{k+1},x_{k+2})\dotsb B^*(x_{k+2M-1},x_{k+2M})
\\
&\qquad\qquad
\times\operatorname{Symm}_{x_1\dots x_{k+2M}}
\bigl(b(x_1,s_1)\dotsb b(x_k,s_k)
\\
&\qquad\qquad\qquad\qquad
\times
 B(x_{k+1},x_{k+2})\dotsb B(x_{k+2M-1},x_{k+2M})\bigr)
\biggr\}
\\
&\qquad\qquad
\times\biggl\{\sum^{\infty}_{k'=0}\sum^{\infty}_{M'=0}
\frac{1}{k'!\mspace{1mu}M'!}
\sum^{\infty}_{s'_1=0}\dots\sum^{\infty}_{s'_{k'}=0}
\int\dots\int dz_1\dots dz_{k'+2M'}
\\
&\qquad\qquad
\times b^*(z_1,s'_1)\dotsb b^*(z_{k'},s'_{k'})
 B^*(z_{k'+1},z_{k'+2})\dotsb B^*(z_{k'+2M'-1},z_{k'+2M'})
\\
&\qquad\qquad
\times\operatorname{Symm}_{z_1\dots z_{k'+2M'}}
\bigl(b(z_1,s'_1)\dotsb b^*(z_{k'},s'_{k'})
\\[-1mm]
&\qquad\qquad\qquad\qquad
\times
 B^*(z_{k'+1},z_{k'+2})\dotsb B(z_{k'+2M'-1},z_{k'+2M'})\bigr)
\biggr\}^{-1}.
\end{split}
\end{equation}

 In the corresponding fermionic formula,
$\operatorname{Symm}$
is replaced by~$\operatorname{Asymm}$.

\section{SYMBOL OF AN ULTRASECOND-QUANTIZED OPERATOR}
\label{ss4}
 First of all, note that the above definition
of pseudosymbol does not fully reflect the
thermodynamic asymptotics, even though it complies
with the Bogolyubov--Dirac rule saying that
the creation--annihilation operators in the leading
asymptotic term should be replaced by $c$-numbers.
 In any case, we refer to the operator thus defined
as the pseudosymbol.
 Let the operator~$\widehat H$
have the form
\begin{equation}
\begin{split}
\label{eq23}
\widehat{H}&=\sum_{l=1}^{L}
\int\dots\int dx_1\dots dx_l\,\,
\widehat{\psi}^+(x_1)\dotsb\widehat{\psi}^+(x_l)
\\
&\qquad\qquad
\times H_l\biggl(\stackrel{2}{x_1},\dots,
\stackrel{2}{x_l};
-i\frac{\stackrel{1}{\partial}}{\partial x_1}\mspace{1mu},
\dots,-i\frac{\stackrel{1}{\partial}}{\partial x_l}\biggr)
\widehat{\psi}^-(x_l)\dotsb\widehat{\psi}^-(x_1).
\end{split}
\end{equation}
 Then, in the case of ultrasecond quantization
without creation--annihilation operators~$\widehat B^\pm(x,y)$
for pairs of particles,
the operators~$\overline{\widehat H}$
and~$\overline{\widehat E}$ defined above satisfy
the identity
\begin{equation}
\label{eq24}
\overline{\widehat{H}}
=\overline{\widehat{E}}\,
\overline{\widehat{A}},
\end{equation}
where~$\overline{\widehat A}$ is an operator in~$\mathscr F$
of the form
\begin{equation}
\begin{split}
\label{eq25}
\overline{\widehat{A}}
&=\sum_{l=1}^{L}\sum_{s_1=0}^\infty\dots
\sum_{s_l=0}^\infty\int\dots\int dx_1\dots dx_l\,\,
\widehat{b}^+(x_1,s_1)\dotsb\widehat{\psi}^+(x_l,s_l)
\\
&\qquad\qquad
\times H_l\biggl(\stackrel{2}{x_1},\dots,
\stackrel{2}{x_l};
-i\frac{\stackrel{1}{\partial}}{\partial x_1}\mspace{1mu},
\dots,-i\frac{\stackrel{1}{\partial}}{\partial x_l}\biggr)
\widehat{b}^-(x_l,s_l)\dotsb\widehat{b}^-(x_1).
\end{split}
\end{equation}

 If ultrasecond quantization also takes into account
creation--annihilation operators
for pairs of particles, then identity~\eqref{eq24} remains valid,
but the operator~$\overline{\widehat A}$
has a more complicated form than~\eqref{eq25}.
 However, for pairs, the operator~$\overline{\widehat A}$
has the form
\begin{equation}
\begin{split}
\label{eq26}
\overline{\widehat{A}}
&=\sum_{s=0}^\infty\int dx\,\,\widehat{b}^+(x,s)
\biggl(-\frac{h^2}{2m}\Delta\biggr)\widehat{b}^-(x,s)
\\
&\qquad
+\iint dx\,dy\,\,\widehat{B}^+(x,y)
\biggl(-\frac{h^2}{2m}(\Delta_x+\Delta_y)\biggr)
\widehat{B}^-(x,y)
\\
&\qquad
+\frac12\sum_{s_1=0}^\infty\sum_{s_2=0}^\infty
\iint dx\,dy\,\,V(x,y)\widehat{b}^+(x,s_1)
\widehat{b}^+(y,s_2)\widehat{b}^-(y,s_2)
\widehat{b}^-(x,s_1)
\\
&\qquad
+\sum_{s=0}^\infty\iiint dx\,dy\,dz\,\,(V(x,y)+V(x,z))
\widehat{b}^+(x,s)\widehat{B}^+(y,z)
\widehat{B}^-(y,z)\widehat{b}^-(x,s)
\\
&\qquad
+\iint dx\,dy\,\,V(x,y)\widehat{B}^+(x,y)
\widehat{B}^-(x,y)
\\
&\qquad
+\frac12\iiiint dx\,dy\,dz\,dw\,\,
 V(x,y)\widehat{B}^+(x,y)\widehat{B}^+(z,w)
\\
&\qquad\qquad
\times\biggl(\widehat{B}^-(y,w)\widehat{B}^-(x,z)
+\widehat{B}^-(w,y)\widehat{B}^-(z,x)
\\
&\qquad\qquad\qquad
+\widehat{B}^-(y,z)\widehat{B}^-(w,x)
+\widehat{B}^-(z,y)\widehat{B}^-(x,w)\biggr).
\end{split}
\end{equation}

 If we replace the operators~$\widehat B^\pm(x,y)$
in~\eqref{eq25} and~\eqref{eq26} by~$c$-numbers,
then we obtain the symbol corresponding to the
asymptotics in the thermodynamic limit.

\section{BOSONIC CASE.
 BOGOLYUBOV FORMULA}
\label{ss5}
 Consider a system of~$N$ identical bosons of mass~$m$ in
a three-dimensional parallelepiped~$\mathbf T$
with side lengths~$L_1$, $L_2$, and~$L_2$.
 We assume that bosons interact with each other
and the interaction potential has the form
\begin{equation}
\label{eq27}
 V(N^{1/3}(x-y)),
\end{equation}
where~$V(\xi)$ is a compactly supported even function
and~$x$ and~$y$ are the coordinates of the bosons
on~$\mathbf T$.
 We assume the periodic boundary conditions along~$L_1$
and impose the conditions that the derivatives are zero
along~$L_2$.
 Note that the interaction potential~\eqref{eq27} depends
on~$N$ in such a way that its range decreases with increasing
particle number~$N$, while the mean number of particles interacting
with a given particle remains constant.

 An explicit expression for the ultrasecond-quantized
operator~$\overline{\widehat H}$ corresponding to the
bosonic system in question under ultrasecond quantization
with pairs is given earlier in this paper.
 As was discussed above, this ultrasecond-quantized
operator satisfies the identity
\begin{equation}
\label{eq28}
\overline{\widehat{H}}=\overline{\widehat{E}}\widehat{A},
\end{equation}
where $\overline{\widehat E}$ is the ultrasecond-quantized
identity operator and~$\widehat A$ is an operator in the
ultrasecond quantization space.
 It is easily seen that the operator
\begin{equation}
\begin{split}
\label{eq29}
\widehat{A}&=\iint dx\,dy\,\,\widehat{B}^+(x,y)
\biggl(-\frac{\hbar^2}{2m}(\Delta_x+\Delta_y)
+V(N^{1/3}(x-y))\biggr)\widehat{B}^-(x,y)
\\
&\qquad
+2\iiiint dx\,dy\,dx'\,dy'\,\,
 V(N^{1/3}(x-y))\widehat{B}^+(x,y)
\widehat{B}^+(x',y')\widehat{B}^-(x,x')
\widehat{B}^-(y,y'),
\end{split}
\end{equation}
where~$\widehat B^+(x,y)$
and~$\widehat B^-(x,y)$ are the bosonic creation--annihilation
operators for a pair of particles in the Fock space of ultrasecond
quantization, satisfies identity~\eqref{eq28}.
 By~\eqref{eq28}, to find the asymptotics of the spectrum of the
bosonic system in question in the limit as~$N\to\infty$,
one needs to find the corresponding asymptotics for the
operator~\eqref{eq29}.

 Since the function~\eqref{eq27} multiplied by~$N$
weakly converges as~$N\to\infty$ to the Dirac delta function,
the second term in the operator~\eqref{eq29} has the factor~$1/N$
in this limit case.
 This means that, to find the asymptotics of eigenvalues and
eigenfunctions of~$\widehat A$, one can apply the semiclassical
methods developed in~\cite{9}.
 The asymptotics of eigenvalues and eigenfunctions is determined
by the symbol of the operator~\eqref{eq29}; this symbol is called
the \textit{true symbol} of the ultrasecond-quantized problem.
 The true symbol corresponding to the operator~\eqref{eq29}
is given by the following functional defined for a pair of
functions~$\Phi^+(x,y)$ and~$\Phi(x,y)$:
\begin{equation}
\begin{split}
\mathscr H[\Phi^+(\,\cdot\,),\Phi(\,\cdot\,)]
&=\iint dx\,dy\,\,\Phi^+(x,y)\biggl(-\frac{\hbar^2}{2m}(\Delta_x+\Delta_y)\biggr)\Phi(x,y)
\\
&\qquad
+2\iiiint dx\,dy\,dx'\,dy'\,\,
(NV(N^{1/3}(x-y)))\Phi^+(x,y)
\\
&\qquad\qquad
\times\Phi^+(x',y')
\Phi(x,x')\Phi(y,y').
\label{eq30}
\end{split}
\end{equation}
 Since the number of particles is preserved in the system
for the functions~$\Phi^+(x,y)$
and~$\Phi(x,y)$, we obtain the condition
\begin{equation}
\iint dx\,dy\,\,\Phi^+(x,y)\Phi(x,y)=\frac12\mspace{1mu}.
\label{eq31}
\end{equation}

 By the asymptotic methods in~\cite{9},
to each solution of the system
\begin{equation}
\label{eq32}
\Omega\Phi(x,y)=\frac{\delta\mathscr H}
{\delta\Phi^+(x,y)}\mspace{1mu},\qquad
\Omega\Phi^+(x,y)=\frac{\delta\mathscr H}
{\delta\Phi(x,y)}\mspace{1mu}
\end{equation}
with condition~\eqref{eq31},
there corresponds an asymptotic series of
eigenfunctions and eigenvalues of the operator~\eqref{eq29}
in the limit as~$N\to\infty$.
 It follows from the explicit form of the true symbol~\eqref{eq30}
that system~\eqref{eq32} can be represented in the form
\begin{equation}
\label{eq33}
\begin{aligned}
\Omega\Phi(x,y)
&=-\frac{\hbar^2}{2m}(\Delta_x+\Delta_y)\Phi(x,y)
\\
&\qquad
+\iint dx'\,dy'\,\,\bigl(NV(N^{1/3}(x-y))
+NV(N^{1/3}(x'-y'))\bigr)
\\
&\qquad\qquad
\times\Phi^+(x',y')
\Phi(x,x')\Phi(y,y'),
\\
\Omega\Phi^+(x,y)
&=-\frac{\hbar^2}{2m}(\Delta_x+\Delta_y)\Phi^+(x,y)
\\
&\qquad
+2\iint dx'\,dy'\,\,\bigl(NV(N^{1/3}(x-x'))
+NV(N^{1/3}(y-y'))\bigr)
\\
&\qquad\qquad
\times\Phi(x',y')
\Phi^+(x,x')\Phi^+(y,y').
\end{aligned}
\end{equation}
 Let~$v_q$ be the coefficient in the Fourier series expansion
of the potential~$NV(N^{1/3}x)$
on the parallelepiped~$(L_1,L_2,L_2)$:
\begin{equation}
\label{eq34}
v_q=\frac{1}{L_1L_2^2}\int_Te^{-iqx}
 NV\bigl(\sqrt[3]{N}(x)\bigr)\,dx,\qquad
v_{-q}=v_q.
\end{equation}
 The exact solution of system~\eqref{eq33} is given
by the functions
\begin{align}
\Phi^+_{k_1,k_2}&=\frac{1}{L_1L_2^2}\mspace{1mu}
e^{-ik_1(x+y)}\cos(k_2(x-y)),
\label{eq35}
\\
\Phi_{k_1,k_2}&=\frac{1}{L_1L_2^2}
\sum_l\varphi_{k_2,l}e^{ik_1(x+y)}e^{il(x-y)},
\label{eq36}
\end{align}
where~$k_1=(n_1/L_1,0,0)$ and~$k_2=(0,n_2/L_2,n_3/L_2)$,
with the eigenvalue
\begin{equation}
\label{eq37}
\Omega=\frac{h^2}{m}
(k_1^2+k_2^2)+v_0+v_{2k_2},
\end{equation}
where the function~$\varphi_{k_2,l}$ has the form
\begin{equation}
\label{eq38}
\begin{gathered}
\begin{aligned}
\varphi_{k_2,l}&=-\frac{b_l}{2}
+\frac12\sqrt{b^2_l-1}\mspace{1mu},\qquad
l^2>k_2^2,
\\
\varphi_{k_2,l}&=-\frac{b_l}{2}
-\frac12\sqrt{b^2_l-1}\mspace{1mu},\qquad
l^2<k_2^2,
\end{aligned}
\\
\varphi_{k_2,k_2}=\frac12\mspace{1mu},\qquad
\varphi_{k_2,l}=\varphi_{k_2,-l},
\\[1mm]
b_l=\frac{h^2/m(l^2-k^2_2)-(v_0+v_{2k_2})}
{v_{l-k_2}+v_{l+k_2}}\mspace{1mu},\qquad
b_l=b_{-l}.
\end{gathered}
\end{equation}
(If~$v_l\to0$, then~$\varphi_{k_2,l}\to0$.)

 The pair $(k_1,k_2)$ of vectors plays the role of parameters
numbering various solutions of this system.
 The vector~$\hbar k_1/m$ is equal to the bosonic system
flow velocity along the capillary.
 The vector~$k_2$ is the wave vector of the transverse mode.

 Note that~$b_l\to\infty$ as~$|l|\to\infty$,
since
\begin{equation}
\label{eq39}
|v_l|=\frac{1}{L_1L_2^2}
\int_{NT}e^{-il\xi/N}V(\xi)\,d\xi
\le\frac{1}{L_1L_2^2}\int_{NT}|V(\xi)|\,d\xi
<\frac{1}{L_1L_2^2}\int_{R^3}|V(\xi)|\,d\xi,
\end{equation}
and consequently
\begin{equation}
\label{eq40}
\varphi_{k,l}\cong\frac{1}{b_l^2}\mspace{1mu},
\end{equation}
whence it follows that the series~\eqref{eq36}
converges absolutely.

 We split the series~\eqref{eq36} into two parts,
one with~$l\le N^{1/6}$ and the other with~$l>N^{1/6}$.
 The first part of the sum converges as~$N\to\infty$
modulo~$O(N^{-1/6})$ to
\begin{equation}
\begin{aligned}
\label{eq41}
b_l&\to\frac{h^2(l^2-k^2_2)}{2mV_0}
-1\stackrel{\mathrm{def}}{=}b^0_l,
\\
\varphi_{k_2,l}&\to-\frac{b_l}{2}
\pm\frac12\sqrt{b^2_l-1}\stackrel{\mathrm{def}}{=}
\varphi^0_{k_2,l}.
\end{aligned}
\end{equation}
 This readily follows from the change of
variables~$\sqrt[3]Nx=\xi$ in~\eqref{eq34}.

 The second part of the sum converges to zero
as~$O(N^{-1/6})$ by~\eqref{eq40}.
 Hence system~\eqref{eq33} supplemented with
condition~\eqref{eq31} has the following family of
solutions for~$k_1=0$ in the limit as~$N\to\infty$:
\begin{equation}
\label{eq42}
\begin{aligned}
\Phi^+_{k}(x,y)&=\frac1{L_1L_2^2}\cos(k(x-y)),
\\
\Phi_{k}(x,y)&=\frac1{L_1L_2^2}
\sum_{l}\varphi_{k,l}\exp(il(x-y)),
\end{aligned}
\end{equation}
where~$k$ and~$l$ are three-dimensional vectors of the form
$$
2\pi\biggl(0,\frac{n_2}{L_2}\mspace{1mu},
\frac{n_3}{L_2}\biggr),
$$
$n_2$ and~$n_3$ are integers,
the~$\varphi_{k,l}$ in~\eqref{eq28} have the form
\begin{equation}
\label{eq43}
\varphi^0_{k_2,l}=\frac1{2V_0}\biggl(\frac{\hbar^2}{2m}(k_2^2-l^2)
+V_0\pm\sqrt{\biggl(\frac{\hbar^2}{2m}(k_2^2-l^2)+V_0\biggr)^2
-V_0^2}\biggr),
\end{equation}
(the plus sign is taken for~$l^2>k_2^2$,
and the minus sign, for~$l^2<k_2^2$),
and finally~$V_0$ stands for the expression
\begin{equation}
 V_0=\frac1{L_1L_2^2}\int dx\,V(x),
\label{eq44}
\end{equation}
the integral being taken over~$\mathbf R^3$.
 The vector~$k$ in~\eqref{eq42} plays the role of a parameter
numbering various solutions of system~\eqref{eq33},~\eqref{eq31}.
 The solutions~\eqref{eq42} are standing waves;
there is no flow in the corresponding series.

 The leading asymptotic term of the eigenvalues for the series
corresponding to the solution~\eqref{eq35},~\eqref{eq36} is equal
to~$N$ times the value of the symbol~\eqref{eq30} on these functions:
\begin{equation}
\label{eq45}
 E_{k_1,k_2}=N\biggl(\frac{\hbar^2(k_1^2+k_2^2)}{2m}
+\frac{V_{0}}{2}\biggr).
\end{equation}

 The asymptotics of eigenvalues and eigenfunctions (in particular,
the terms following~$E_{k_1,k_2}$) is determined not only by
system~\eqref{eq33} but also by the solutions of the variational
system corresponding to the Hamiltonian system.
 The variational system for~\eqref{eq33} has the form
\begin{equation}
\label{eq46}
\begin{aligned}
(\Omega-\lambda)F(x,y)
&=-\frac{\hbar^2}{2m}(\Delta_x+\Delta_y)F(x,y)
\\
&\qquad
+2N\iint dx'\,dy'\,\,\bigl(V(\sqrt[3]{N}(x-y))
+V(\sqrt[3]{N}(x'-y'))\bigr)
\\
&\qquad\qquad
\times\bigl(G(x',y')\Phi(x,x')\Phi(y,y')
+\Phi^{+}(x',y')F(x,x')\Phi(y,y')
\\
&\qquad\qquad\qquad
+\Phi^{+}(x',y')\Phi(x,x')F(y,y')\bigr),
\\
(\Omega+\lambda)G(x,y)
&=-\frac{\hbar^2}{2m}(\Delta_x+\Delta_y)G(x,y)
\\
&+2N\iint dx'\,dy'\,\,\bigl(V(\sqrt[3]{N}(x-x'))
+V(\sqrt[3]{N}(y-y'))\bigr)
\\
&\qquad\qquad
\times\bigl(F(x',y')\Phi^{+}(x,x')\Phi^{+}(y,y')
+\Phi(x',y')G(x,x')\Phi^{+}(y,y')
\\
&\qquad\qquad\qquad
+\Phi(x',y')\Phi^{+}(x,x')G(y,y')\bigr).
\end{aligned}
\end{equation}

 To find the spectrum of quasiparticles, one should select
solutions of the variational system satisfying the selection
rule~\cite{10} for the complex germ for self-adjoint operators
with real spectrum.

 Consider the case in which~$k_2\ne0$.
 We substitute the solutions~\eqref{eq35},~\eqref{eq36}
into~\eqref{eq46} and take into account symmetry.
 Then we see that the solutions of the variational system
have the form
\begin{equation}
\label{eq47}
\begin{aligned}
 G_{l}(x,y)&=u_{1,l}\bigl(\exp(i(-k_1+k_2)x+i(-k_1+l)y)
+\exp(i(-k_1+k_2)y+i(-k_1+l)x)\bigr)
\\
&\qquad
+u_{2,l}\bigl(\exp(i(-k_1-k_2)x+i(-k_1+2k_2+l)y)
\\
&\qquad\qquad\qquad
+\exp(i(-k_1-k_2)y+i(-k_1+2k_2+l)x)\bigr),
\\
 F_{l}(x,y)&=-v_{1,l}\bigl(\exp(i(k_1+k_2)x+i(k_1+l)y)
+\exp(i(k_1+k_2)y+i(k_1+l)x)\bigr)
\\
&\qquad
-v_{2,l}\bigl(\exp(i(k_1-k_2)x+i(k_1+2k_2+l)y)
\\
&\qquad\qquad\qquad
+\exp(i(k_1-k_2)y+i(k_1+2k_2+l)x)\bigr)
\\
&\qquad
+\sum_{l'\ne l,l+2k_2}w_{l,l'}\bigl(\exp(i(k_1+k_2+l-l')x+i(k_1+l')y)
\\
&\qquad\qquad\qquad
+\exp(i(k_1+k_2+l-l')y+i(k_1+l')x)\bigr),
\end{aligned}
\end{equation}
where~$l\ne-k_2$ and the numerical coefficients~$u_{1,l}$,
$u_{2,l}$, $v_{1,l}$, $v_{2,l}$, and~$w_{l,l'}$ are determined from
an infinite system of equations.
 This system contains a closed subsystem of four equations
for the coefficients~$u_{1,l}$, $u_{2,l}$, $v_{1,l}$, and~$v_{2,l}$,
which can be rewritten in the form
\begin{equation}
\label{eq48}
\widetilde{\lambda}X=MX,
\end{equation}
where
$$
\widetilde\lambda
=\lambda+\frac{\hbar^2}m\mspace{1mu}k_1(k_2+l)
$$
and $X$~is the column vector
$$
X=\begin{pmatrix}
u_{1,l}
\\
u_{2,l}
\\
v_{1,l}
\\
v_{2,l}
\end{pmatrix}.
$$

 Equations~\eqref{eq47} and~\eqref{eq48} give the matrix
$M$
{\small
$$
M=\begin{pmatrix}
B_l+\dfrac{v_{l-k_2}}2
&\dfrac{v_{2k_2}+v_{l+k_2}}2
&-\dfrac{v_{l+k_2}+v_{l-k_2}}2 &0
\\[3mm]
\dfrac{v_{2k_2}+v_{l+k_2}}2
&B_{l+2k_2}+\dfrac{v_{l+3k_2}}2
&0 &-\dfrac{v_{l+k_2}+v_{l+3k_2}}2
\\[3mm]
2(v_0+v_{l-k_2})\varphi_{k_2,l}
&\mspace{5mu}\substack{\displaystyle\mspace{-63mu}(v_{2k_2}+v_{l+k_2})
\\
\displaystyle
\times(\varphi_{k_2,l}+\varphi_{k_2,l+2k_2})}
&-B_l-\dfrac{v_{l-k_2}}2
&-\dfrac{v_{2k_2}+v_{l+k_2}}2
\\[3mm]
\mspace{5mu}\substack{\displaystyle\mspace{-63mu}(v_{2k_2}+v_{l+k_2})
\\
\displaystyle
\times(\varphi_{k_2,l}+\varphi_{k_2,l+2k_2})}
&2(v_0+v_{l+3k_2})\varphi_{k_2,l+2k_2}
&-\dfrac{v_{2k_2}+v_{l+k_2}}2
&-B_{l+2k_2}-\dfrac{v_{l+3k_2}}2
\end{pmatrix},
$$}
where the~$B_l$ have the form
$$
B_l=\frac{\hbar^2}{2m}(l^2-k_2^2)
+(v_{l-k_2}+v_{l+k_2})\varphi_{k_2,l}
-\frac{v_{2k_2}}2\mspace{1mu}.
$$
 Obviously,
\begin{equation}
\label{eq49}
v_{l+k_2}+v_{l+3k_2}
=v_{l-k_2}+v_{l+k_2}+O\biggl(\frac1N\biggr)
\end{equation}
uniformly in~$l$ as~$N\to\infty$ (it suffices to make
the change of variables~\eqref{eq39}).
 Then the matrix~$M$ can be approximately represented
as the block matrix
$$
M=\begin{pmatrix}
C &-V_lE
\\
D &-C
\end{pmatrix},
$$
where $E$ is the identity~$2\times2$ matrix and
$$
V_l=\frac{v_{l-k_2}+v_{l+k_2}}2\mspace{1mu}.
$$

 We also introduce
$$
V_{l}^{+}=\frac{v_{l+k_2}+v_{2k_2}}{2}\mspace{1mu},\qquad
V_{l}^{-}=\frac{v_{l-k_2}+v_{0}}{2}\mspace{1mu}.
$$

 The eigenvalue corresponding to Eq.~\eqref{eq48} have the form
\begin{equation}
\label{eq50}
\begin{split}
\lambda_{k_1,k_2,l}&=-2ak_1(k_2+l)
\\
&\qquad
\pm\biggl(\frac{1}{2}(a(l^2-k_2^2)+V_{l}-V_{l}^{+})^2
+\frac{1}{2}(a(l_1^2-k_2^2)+V_{l}-V_{l}^{+})^2
+{V_{l}^{+}}^{2}-V_{l}^2
\\
&\qquad
\pm\frac{1}{2}
(a(l_1^2+l^2-2k_2^2)+2V_{l}-2V_{l}^{+})
\sqrt{a^2(l_1^2-l^2)^2+4{V_{l}^{+}}^{2}}\,\biggr)^{1/2},
\end{split}
\end{equation}
where
$$
a=\frac{\hbar^2}{2m}\mspace{1mu},\qquad
l_1=l+2k_2.
$$

 Before passing to the limit, one has
\begin{gather*}
\lambda_{k_1,k_2,l}
=-2ak_1l\pm\biggl((al^2+V_{l}-V_{l}^{+})^2
+{V_{l}^{+}}^{2}-V_{l}^2
\pm2(al^2+V_{l}-V_{l}^{+})|V_{l}^{+}|\biggr)^{1/2},
\\
 V_l=v_l,\qquad
 V_l^{+}=V_l^{-}=\frac{v_l+v_0}{2}\mspace{1mu}
\end{gather*}
for $k_2=0$.
 By formally setting~$k_2=0$, we arrive at Bogolyubov's well-known
formula
$$
\lambda_{1,l}=-\frac{\hbar^2}m\mspace{1mu}k_1l
+\sqrt{\biggl(\frac{\hbar^2l^2}{2m}+v_l\biggr)^2-v_l^2}\,.
$$
 Here~$v_l$ is the Fourier transform of the potential.

(We assume that~$L_2$ is much larger than some standard length,
say, the electron radius~$r_0$, and that although~$L_1\gg L_2$,
we can take a sufficiently large integer~$n_1$.
 In other words, $L_2/r_0\to\infty$ and~$L_1/r_0\to\infty$,
but the vector~$k_1=(n_1/L_1,0,0)$
remains finite, since~$n_1\to\infty$.

 In the language of nonstandard analysis, this means
that~$L_2$ is an infinite (nonstandard) number,
$L_1$ and~$n_1$ are nonstandard numbers of higher order,
and~$k_1=(n_1/L_1,0,0)$ is a standard finite number.

 Then~$k_2$ is equal to an infinitesimal nonstandard zero,
$k_2\cong0$, and~$k_1$ is a standard number.)

\section{CASE OF A FERMI FLUID}
\label{ss6}
Consider the Hamiltonian system for fermions:
\begin{equation}
\label{eq51}
\begin{aligned}
\Omega\Phi(x,y)&=\biggl(-\frac{\hbar^2}{2m}
(\Delta_x+\Delta_y)\biggr)\Phi(x,y)
\\
&\qquad
+2N\iint dx'\,dy'\,\,(V(x-y)+V(x'-y'))\Phi^{+}(x',y')
\Phi(x,x')\Phi(y',y),
\\
\Omega\Phi^{+}(x,y)&=\biggl(-\frac{\hbar^2}{2m}
(\Delta_x+\Delta_y)\biggr)\Phi^{+}(x,y)
\\
&\qquad
+2N\iint dx'\,dy'\,\,(V(x-x')+V(y-y'))\Phi(x',y')
\Phi^{+}(x,x')\Phi^{+}(y',y).
\end{aligned}
\end{equation}
 The functions~$\Phi^+(x,y)$ and~$\Phi(x,y)$ are antisymmetric
and satisfy the normalization condition
\begin{equation}
\label{eq52}
\iint dx\,dy\,\,\Phi^{+}(x,y)\Phi(x,y)
=\frac12\mspace{1mu}.
\end{equation}
 Let us represent the interaction potential by a Fourier series:
$$
NV(x)=\sum_pv_pe^{ipx},\qquad
v_p=\frac1{L_1L_2^2}\int dx\,\,
NV(x)e^{-ipx},\quad
v_p=v_{-p}.
$$

 We seek the solution of system~\eqref{eq51},~\eqref{eq52}
in the form
\begin{equation}
\label{eq53}
\begin{aligned}
\Phi^{+}_{k_1,k_2}(x,y)
&=\frac{1}{L_1L_2^2}\mspace{1mu}
e^{-ik_1(x+y)}\sin(k_2(x-y)),
\\
\Phi_{k_1,k_2}(x,y)
&=\frac{1}{L_1L_2^2}\sum_{l}\varphi_{k_2,l}
e^{il(x-y)+ik_1(x+y)},
\end{aligned}
\end{equation}
where~$k_1$, $k_2$, and~$l$ are three-dimensional vectors of the form
$$
2\pi\biggl(\frac{n_1}{L_1}\mspace{1mu},
\frac{n_2}{L_2}\mspace{1mu},\frac{n_3}{L_2}\biggr)
$$
and~$n_1$, $n_2$, and~$n_3$ are integers.
 The numbers~$\varphi_{k_2,l}$ should satisfy the condition
$$
\varphi_{k_2,l}=-\varphi_{k_2,-l}.
$$
 After the substitution, we find that the eigenvalue is equal to
$$
\Omega=\frac{\hbar^2}m(k_1^2+k_2^2)+v_{2k_2}-v_0
$$
and the~$\varphi_{k_2,l}$ have the form
$$
\varphi_{k_2,l}=-\frac{ib_{k_2,l}}{2}
\pm\frac{1}{2}\sqrt{1-b_{k_2,l}^2}\,,\qquad
b_{k_2,l}\equiv\frac{(\hbar^2/m)(l^2-k_2^2)
+(v_{0}-v_{2k_2})}{v_{l-k_2}-v_{l+k_2}}\mspace{1mu}.
$$
 Note that
$$
b_{k_2,l}=-b_{k_2,-l}.
$$
 Set
$$
\varphi_{k_2,l}=-\frac{ib_{k_2,l}}2
+\frac12\frac{v_{l-k_2}-v_{l+k_2}}
{|v_{l-k_2}-v_{l+k_2}|}\sqrt{1-b_{k_2,l}^2}\,.
$$
 Then~$\varphi_{k_2,l}$ will be equal to~$-\varphi_{k_2,-l}$.

 Consider the fermionic variational system
\begin{equation}
\label{eq54}
\begin{aligned}
(\Omega-\lambda)F(x,y)
&=-\frac{\hbar^2}{2m}(\Delta_x+\Delta_y)F(x,y)
\\
&\qquad
+2N\iint dx'\,dy'\,\,\bigl(V(\sqrt[3]{N}(x-y))
+V(\sqrt[3]{N}(x'-y'))\bigr)
\\
&\qquad\qquad
\times\bigl(G(x',y')\Phi(x,x')\Phi(y',y)
+\Phi^{+}(x',y')F(x,x')\Phi(y',y)
\\
&\qquad\qquad\qquad
+\Phi^{+}(x',y')\Phi(x,x')F(y',y)\bigr),
\\
(\Omega+\lambda)G(x,y)
&=-\frac{\hbar^2}{2m}(\Delta_x+\Delta_y)G(x,y)
\\
&\qquad
+2N\iint dx'\,dy'\,\,\bigl(V(\sqrt[3]{N}(x-x'))
+V(\sqrt[3]{N}(y-y'))\bigr)
\\
&\qquad\qquad
\times\bigl(F(x',y')\Phi^{+}(x,x')\Phi^{+}(y',y)
+\Phi(x',y')G(x,x')\Phi^{+}(y',y)
\\
&\qquad\qquad\qquad
+\Phi(x',y')\Phi^{+}(x,x')G(y',y)\bigr).
\end{aligned}
\end{equation}

 Its solutions have the form
\begin{equation}
\label{eq55}
\begin{aligned}
 G_{l}(x,y)&=u_{1,l}
(e^{i(-k_1+k_2)x+i(-k_1+l)y}
-e^{i(-k_1+k_2)y+i(-k_1+l)x})
\\
&\qquad
+u_{2,l}(e^{i(-k_1-k_2)x+i(-k_1+2k_2+l)y}
-e^{i(-k_1-k_2)y+i(-k_1+2k_2+l)x}),
\\
 F_{l}(x,y)&=v_{1,l}
(e^{i(k_1+k_2)x+i(k_1+l)y}
-e^{i(k_1+k_2)y+i(k_1+l)x})
\\
&\qquad
+v_{2,l}(e^{i(k_1-k_2)x+i(k_1+2k_2+l)y}
-e^{i(k_1-k_2)y+i(k_1+2k_2+l)x})
\\
&\qquad
+\sum_{l'\ne l,l+2k_2}w_{l,l'}
(e^{i(k_1+k_2+l-l')x+i(k_1+l')y}
-e^{i(k_1+k_2+l-l')y+i(k_1+l')x}),
\end{aligned}
\end{equation}
where~$l\ne-k_2$ and the numerical coefficients~$u_{1,l}$,
$u_{2,l}$, $v_{1,l}$, $v_{2,l}$, and~$w_{l,l'}$ are determined
from an infinite system of equations.
 This system contains a closed subsystem of four equations
for the coefficients~$u_{1,l}$, $u_{2,l}$, $v_{1,l}$,
and~$v_{2,l}$, which can be rewritten in the
standard form~\eqref{eq48}.

The matrix~$M$ is given by
{\small
$$
M=\begin{pmatrix}
B_l+\dfrac{v_{l-k_2}}2 &\dfrac{v_{l+k_2}-v_{2k_2}}2
&\dfrac{v_{l-k_2}-v_{l+k_2}}2 &0
\\[3mm]
\dfrac{v_{l+k_2}-v_{2k_2}}2
&B_{l+2k_2}+\dfrac{v_{l+3k_2}}2
&0 &\dfrac{v_{l+3k_2}-v_{l+k_2}}2
\\[3mm]
2i(v_{l-k_2}-v_0)\varphi_{k_2,l}
&\mspace{8mu}\substack{\displaystyle\mspace{-63mu}
i(v_{2k_2}-v_{l+k_2})
\\
\displaystyle
\times(\varphi_{k_2,l+2k_2}-\varphi_{k_2,l})}
&-B_l-\dfrac{v_{l-k_2}}2
&-\dfrac{v_{l+k_2}-v_{2k_2}}2
\\[3mm]
\mspace{8mu}\substack{\displaystyle\mspace{-63mu}
i(v_{2k_2}-v_{l+k_2})
\\
\displaystyle
\times(\varphi_{k_2,l+2k_2}-\varphi_{k_2,l})}
&2i(v_0-v_{l+3k_2})\varphi_{k_2,l+2k_2}
&-\dfrac{v_{l+k_2}-v_{2k_2}}2
&-B_{l+2k_2}-\dfrac{v_{l+3k_2}}2
\end{pmatrix}\!,
$$}
and the~$B_l$ have the form
$$
B_l=\frac{\hbar^2}{2m}(l^2-k_2^2)
+i(v_{l+k_2}-v_{l-k_2})\varphi_{k_2,l}
-\frac{v_{2k_2}}2\mspace{1mu}.
$$

 For~$k_2=0$, the eigenvalues have the form
$$
\lambda_{1,l}=-\frac{\hbar^2}{m}\mspace{1mu}lk_1
+\frac{\hbar^2l^2}{2m}\mspace{1mu},\qquad
\lambda_{2,l}=-\frac{\hbar^2}{m}\mspace{1mu}lk_1
+\biggl|\frac{\hbar^2l^2}{2m}+v_l-v_0\biggr|.
$$

 This means that we choose the vectors~$k_2=(0,1/L_2,0)$
and~$k_1=(n/L_1,0,0)$.
 If~$n\gg1$ and hence~$k_1\gg k_2$,
then the fluid velocity vector is mainly directed
along the tube.
 As before, let~$L_1=\infty$, $L_2=\infty$, and~$n=\infty$
be nonstandard numbers, let~$L_1\gg L_2$, and let~$k_1$
be a standard number.
 Then~$k_2\cong0$, and~$\varphi_{k_2,l}$ takes nonstandard values
for~$k_2\cong0$.
 However, if $0\le l\le\sqrt{k_2}M$, where~$M\le\infty$
(i.e., $0\le l\le\infty$), then~$\varphi_{k_2,l}$ can be assumed to
be a standard number.
 For~$k_2\cong0$,
taking into account the selection rule, we obtain the formula
$$
\lambda_{2,l}=-\frac{\hbar^2}m\mspace{1mu}lk_1
+\biggl|\frac{\hbar^2l^2}{2m}+v_l-v_0\biggr|
$$
by analogy with the bosonic case and formula~\eqref{eq1}.

 As was mentioned in Sec.~\ref{ss2},
the behavior of the expression under the modulus sign
is similar to that of the Landau curve for bosons.
 The criterion for~$k_1$ has the form
$$
|k_1|\le\frac m{\hbar^2}\min_l
\biggl|\frac{v_l-v_0}{|l|}+\frac{\hbar^2l}{2m}\biggr|,
$$
similar to the Landau criterion (the vapor-destroying velocity).
 We set the parameter~$k_2$ to be an infinitesimal
nonstandard number.
 However, it is not exactly zero, and the presence of
a nonzero~$k_2$ results in a spectral gap.
 To compute the gap in the first approximation,
one has to find the spectrum of the matrix~$M$
modulo~$O(k_2^2)$.
 Then, by using the selection rules, one can
determine whether there is a gap in the spectrum.

\section*{ACKNOWLEDGMENTS}

The author wishes to express gratitude to D.~S.~Golikov
for re-calculation and verification of all formulas.

\end{document}